\documentclass{ws-ppl}

\usepackage{graphicx}
\usepackage{multirow}
\begin{document}

\markboth{Parallel Processing Letters}
{An Agent-Based Approach for Fault Tolerance}

%
\catchline{}{}{}{}{}
%

\title{CAN AGENT INTELLIGENCE BE USED TO ACHIEVE FAULT TOLERANT PARALLEL COMPUTING SYSTEMS?}

\author{BLESSON VARGHESE\footnote{Blesson Varghese is a PhD candidate at the School of Systems Engineering (Corresponding Author).}}
\address{School of Systems Engineering, University of Reading, Whiteknights\\
Reading, Berkshire, United Kingdom, RG6 6AY\\
E-mail: b.varghese@pgr.reading.ac.uk}

\author{GERARD MCKEE\footnote{Dr Gerard McKee was formerly a Senior Lecturer in Networked Robotics at the School of Systems Engineering, University of Reading, UK. He is now the Dean of the Faculty of Computing and IT at Baze University, Nigeria.}}
\address{Faculty of Computing and IT, Baze University, Abuja, Nigeria\\
E-mail: gerard.mckee@bazeuniversity.edu.ng}

\author{VASSIL ALEXANDROV\footnote{Prof Vassil Alexandrov is ICREA Research Professor in Computational Science at the Barcelona Supercomputing Centre.}}
\address{Barcelona Supercomputing Centre, C/ Jordi Girona, 29\\Edifici Nexus II, E-08034 Barcelona, Spain\\
E-mail: vassil.alexandrov@bsc.es}

\maketitle

\begin{history}
\received{July 2010}
\revised{August 2010}
\comby{Ke Qiu}
\end{history}

\begin{abstract}
The work reported in this paper is motivated towards validating an alternative approach for fault tolerance over traditional methods like checkpointing that constrain efficacious fault tolerance. \emph{\textbf{Can agent intelligence be used to achieve fault tolerant parallel computing systems?}} If so, \emph{``What agent capabilities are required for fault tolerance?''}, \emph{``What parallel computational tasks can benefit from such agent capabilities?''} and \emph{``How can agent capabilities be implemented for fault tolerance?''} need to be addressed. Cognitive capabilities essential for achieving fault tolerance through agents are considered. Parallel reduction algorithms are identified as a class of algorithms that can benefit from cognitive agent capabilities. The Message Passing Interface is utilized for implementing an intelligent agent based approach. Preliminary results obtained from the experiments validate the feasibility of an agent based approach for achieving fault tolerance in parallel computing systems.
\end{abstract}


\section{Introduction}
Fault tolerance for computing systems is a classic area of research that has been explored by computing researchers over decades. To address fault tolerance, checkpointing techniques have been implemented for traditional computing systems of less complexity. These techniques essentially aim to periodically or non-periodically, when requested or automatically, save on memory or disk the state of execution, and further use the saved information for restarting execution when a failure occurs.

However, since complexity of computing systems have significantly increased in due time, the drawbacks of checkpointing have posed constraints on effectively achieving fault tolerance for such large scale systems. Hence, it has been necessary to develop techniques that can replace traditional checkpointing methods.

Middleware layers have been implemented in many large scale computing system architectures with the aim of addressing the drawbacks of checkpointing. This effort has been able to surmount a few challenges, though hasn't proved much effective.

Distributed artificial intelligence in the form of multi-agents is another plausible technique that can be employed for achieving fault tolerance in computing systems. However, it is noted that there has been less effort towards extending and implementing such ideas for large scale parallel computing systems.

The multi-agent technology is advantageous over traditional fault tolerance due to three reasons. Firstly, since agents in multi-agent systems are characterized by persistence, hence seamless execution of a task in a parallel distributed system without cold restarts is possible. Secondly, agents can contribute effectively towards self-management, hence reduces management responsibilities on an administrator. Thirdly, agents can be mobile over distributed nodes, hence paving way for process migration and cuts down voluminous checkpointing.

Hence, with the above advantages in view, three questions applicable to large scale parallel computing systems, ``What agent capabilities are required for fault tolerance?'', ``What parallel computational tasks can benefit from such agent capabilities?'' and ``How can agent capabilities be implemented for fault tolerance?'' are addressed in this paper. To this end, an agent-based method for achieving fault tolerance is proposed in which a task to be executed on a parallel computing system is decomposed into sub-tasks and mapped onto agents that carry these tasks onto nodes or cores for execution. Cognitive capabilities required for fault tolerance, parallel reduction algorithms that can benefit from such agent capabilities and a computer cluster-based implementation of agent capabilities for improving fault tolerance are presented in this paper as a preliminary effort towards realising the proposed agent-based method.

The remainder of this paper is organised as follows. Section 2 presents the background and related work that has motivated the work reported in this paper. Section 3 considers ``What agent capabilities are required for fault tolerance?''. Section 4 identifies ``What parallel computational tasks can benefit from such agent capabilities?''. Section 5 presents ``How agent capabilities can be implemented for fault tolerance?'', and addresses the question by performing an experiment which is used to gather statistics to evaluate the approach. Section 6 concludes this paper by considering future work.

\section{Background \& Related Work}

With the increase in complexity of computing systems, traditional checkpointing methods have posed constraints on achieving fault tolerance in such large scale systems. There are five major drawbacks that impair checkpointing in being effective for large scale computing systems.

Firstly, server based checkpointing strategies are subject to single point of failure \cite{Replication based Fault Tolerance for MPI Applications}. To address this issue, multi-server checkpointing strategies have been introduced \cite{Replication based Fault Tolerance for MPI Applications}. However, these centralized server strategies tend to be less scalable on complex and heterogeneous environments.

Secondly, checkpointing relies on network storage or shared memories accessible to an entire distributed system, thereby increasing challenges like reliability, scalability and stability of the fault tolerance mechanism in the computing system.

Thirdly, an attempt to checkpoint a large process involves large overheads and greater time to write the checkpoint to a stable storage system. In order to mitigate this issue, distributed commit protocols \cite{Design Implementation and Performance of Fault-Tolerant Message Passing Interface} and diskless checkpointing \cite{Algorithm-based Fault Tolerance for Fail-Stop Failures} strategies based on memory and processor redundancy have been developed. These strategies tend to be ineffective if the checkpoint size and the number of nodes in the distributed system is large.

Fourthly, most checkpoint strategies require a cold restart, that is, a complete reload of all processes associated with the parallel job \cite{FTPA: Supporting Fault-Tolerant Parallel Computing through Parallel Recomputing}. In this case, processors that did not suffer a failure might also require a reload of the process executing on it.

Fifthly, in mobile agent technology, checkpointing can prevent the loss of an agent and prevent blocking. In this case single failure does not prevent the progress of a mobile agent execution. However, checkpointing does not satisfy the exactly-once property, leading to multiple executions of an agent \cite{A New Approach for Mobile Agent Fault-Tolerance and Reliability}.

On an implementation level, checkpointing based fault tolerance has opened avenues for implementing middleware approaches that aim to add an additional interface or a sandwich layer between hardware and software layers \cite{The Robust Middleware Approach for Transparent and Systematic Fault Tolerance in Parallel and Distributed Systems}. To improve efficiency of checkpointing, additional checkpointing strategies over custom implementations have been adopted in such middleware layers like MPI (Message Passing Interface), a few of which are referenced here.

In \cite{Implementing and Evaluating Automatic Checkpointing}, the concept of automatic checkpointing is introduced in LAM/MPI middleware. The strategy records the context of an application periodically, identifies failed nodes and restarts MPI processes only on failed nodes, hence allowing continuity of the executing application by taking advantage of the computing done previously.

In \cite{The Robust Middleware Approach for Transparent and Systematic Fault Tolerance in Parallel and Distributed Systems}, DREAM (Dynamic Robust Embedding/Allocation Middleware) based on Robust MPI (R-MPI) as a library component is proposed. In \cite{Algorithm-based Fault Tolerance for Fail-Stop Failures}, to address challenges in diskless checkpointing, algorithm-based fault tolerance (ABFT) using Fault Tolerant MPI (FT-MPI) is introduced. Recovery from failure in the middle of computations is performed by maintaining a checksum relationship.

In \cite{Replication based Fault Tolerance for MPI Applications}, to address the scalability issue of checkpointing in MPI applications, an asynchronous replication strategy is introduced that distributes replication overhead over all participating nodes in the computation.

In \cite{Design Implementation and Performance of Fault-Tolerant Message Passing Interface}, fault tolerant MPI comprising a replicated system controller, a node controller and checkpoint server tested on a parallelized weather model is introduced. The fault tolerant version is designed to address single point failures, ensure consistency of checkpoint files and robustness of fault detection hierarchy.

In \cite{Fault-Tolerant solutions for a MPI compute Intensive application}, for computationally intensive applications using MPI, two approaches for checkpoint based fault tolerance is proposed. Firstly, segment-level solution, an extension of a checkpoint library for sequential codes. Secondly, variable-level solution, a manual solution determined by the programmer that inserts safe points and specifies data to be stored during checkpointing into program code.

In \cite{A Fault Tolerant Approach in Cluster Computing System}, an extension to MPI is proposed that consists of two steps to achieve fault tolerance. Firstly, failure diagnosis, detection of the location of a failed component. Secondly, failure recovery, a step towards reassigning tasks of a failed component to fully functional system nodes.

However, most of the above solutions implemented in the MPI middleware face similar challenges apparent in traditional checkpointing strategies. To overcome these challenges, in recent times, multi-agent systems have incorporated concepts of fault tolerance. The multi-agent technology is beneficial and has been considered in the previous section.

Research on multi-agent based fault tolerance is reported in \cite{Multiagent Technology or Fault Tolerance and Flexible Control}\cite{Multi-agent Platform for Fault Tolerant Control Systems}\cite{Plan-Based Replication for Fault-Tolerant Multi-Agent Systems}\cite{Decentralized Architecture for Fault Tolerant Multi Agent System}\cite{A Fault-Tolerant Infrastructure for Mobile Agents}. Though research has been pursued on multi-agents focusing on fault tolerance, it is surprising that there has been little effort towards extending and implementing such ideas for large scale parallel computing systems.

To employ multi-agents in large scale parallel computing systems it is necessary to investigate what agent capabilities are required for fault tolerance. Hence, the next section considers the cognitive agent capabilities required to achieve fault tolerance. These capabilities are then further aimed to be implemented in the method proposed in this paper.

\section{Agent Capabilities}
\emph{What agent capabilities are required for fault tolerance?}

Agent-based techniques are biomimetically inspired, that is biologically inspired from nature to foster innovative designs for man-made systems \cite{Biomimetics Robots from Bio-inspiration to Implementation}\cite{Deployment of a team of biomimetic searching agents based on limited communication quantity}. For example, swarming of agents in multi-agent systems like swarm robotic systems are inspired from the biological phenomena of swarming bees.

Agents in a natural swarm also demonstrate intelligence by their cognitive capabilities in at least four different ways. Firstly, an agent is capable of being able to know its environment, the surroundings in which it is located. Secondly, an agent is capable to identify a location in the environment in which it can nicely situate. Thirdly, an agent is capable to sense any hazard that is likely to deteriorate or impair its functioning. Fourthly, an agent is capable to pass over from one location to another when necessary for survival. These capabilities are also desirable for agents in a computing environment. In other words, in a computing environment, cognitive agent intelligence is not demonstrated by merely being able to act as reflexive agents, but by also being able to perceive, reason, judge, respond and learn \cite{Computational Intelligence Based Architecture for Cognitive Agents}.

The aim of the intelligent agent based approach proposed in this paper is to achieve agent intelligence in parallel computing systems and further demonstrate that the cognitive capabilities of an agent complementing its intelligence can lead towards fault tolerance.

In the proposed approach, a task to be executed on a parallel computing system is decomposed into sub-tasks and mapped onto agents that carry these tasks onto nodes or cores for execution. The agent and the sub-problem are independent of each other; in other words, the agents only carry the sub-tasks or act as a wrapper around the sub-task independent of the operations performed by the task.

An agent possesses capabilities similar to the capabilities of a natural agent presented above. Intelligence of an agent in the computing environment is demonstrated in four different ways. Firstly, an agent is aware of its environment, that is the nodes or cores on which it can carry a task onto, other agents in its vicinity and agents with which it interacts or shares information. Secondly, an agent can situate itself on a node or core that may not fail soon and can provide necessary and sufficient consistency in executing the task. Thirdly, an agent can predict core failures by consistent monitoring (for example, power consumption and heat dissipation of the cores can be used to predict failures). Fourthly, an agent is capable of shifting gracefully from one core to another, without causing interruption to the state of execution, and notifying other interacting agents in the system when a core on which a sub-task being executed is predicted to fail.

More specifically, the computing environment in which an agent is situated comprises both other agents with which it can interact and computing resources. Perception in this context would mean to acquire information concerning the environment. For this, an agent needs to answer questions such as `are there other agents in my vicinity?' and `which computing cores are functional in my vicinity?' To achieve this, an agent can probe its environment, i.e., by sending `are you alive' signals to the agent and the computing resources. Perception for an agent also includes gathering information for answering the question `will the core that I am situated on fail?'.

Reasoning for an agent in a computing environment becomes necessary once an agent predicts the computing core it is situated on to fail. An agent needs to answer questions such as `which cores in the computing environment would it be possible to move onto?' Since an agent has options to move onto other cores in its vicinity, an agent needs to make an appropriate choice. Hence, the agent needs to also think `will the core that I will move onto fail?'. For this, an agent should gather sensory information of the cores in its vicinity.

Judging for an agent in a computing environment is necessary for decision making. For example, an agent may think about `which core do I move to?', but a decision has to be made concfirming the core to which an agent can move. As suggested above, the sensory information perceived by an agent aids decision making.

After an agent makes a decision as to which core it can move onto, a response needs to be initiated. A response for example, instructions like `move to' or `move to core x', so that the agents can move onto a core other than which it is situated on is an example.

Learning in the computing environment is based on the perceived sensory information and can also aid decision making. For example, in the context of the algorithm implemented in this section, an agent updates its information on the cores it is dependent on. The core dependencies known to the agent and the knowledge gained from `are you alive' signals contribute to the knowledge of an agent about the computing environment.

\section{Benefitting Tasks}
\emph{What parallel computational tasks can benefit from such agent capabilities?}

One important aspect in large scale parallel computing is binary trees \cite{27}. Algorithms that implement binary trees can have data flow from the leaves of a tree to its root (bottom-up) and is implemented in fan-in or reduction algorithms \cite{27}.

Parallel reduction algorithms, which implement the bottom-up approach of binary trees, are of interest in the context of fault tolerance due to two reasons. Firstly, the computing nodes of a parallel reduction algorithm tend to be critical. The execution of the algorithm stalls or produces an incorrect solution if any node information is lost. Secondly, parallel reduction algorithms are employed in critical applications such as space applications. These applications require fault tolerant distributed systems.

In space applications, processing cores of a Field Programmable Gate Array (FPGA) employed in spacecrafts are subject to Single Event Upsets (SEU) caused by radiation on moving out of earth's atmosphere \cite{36} \cite{37}. If critical applications such as trajectory prediction of a space craft using Kalman filters \cite{38} employ the parallel reduction algorithm without fault tolerance, it is more likely that the execution of the algorithm miscarries. In these cases, neither does checkpointing come into power play due to the drawbacks considered in Section 2. Moreover, such applications do not have the luxury of time to reinstate the cores and restart the execution of algorithms. Hence, a fault-tolerant parallel reduction algorithm are required.

To design such fault tolerant algorithms it is necessary to consider the fault tolerance requirements. First of all, the design must incorporate a reactive or a proactive fault tolerant mechanism. Reactive fault tolerant mechanisms may not prove useful for critical applications since faults that occur within the system are dealt with only after the fault occurs. Proactive fault tolerance, on the other hand, predicts likely faults or failures and takes preventive measures to avoid a likely occurence of a fault. In the case of developing parallel reduction algorithms designed for critical applications it would be appropriate to incorporate proactive fault tolerance.

If proactive fault tolerant is incorporated in the algorithm then what strategy is incorporated to prevent the loss of data also needs to be considered. Data replication on several nodes is one possible strategy. However, if voluminous data needs to replicated then more storage space will be required on computing nodes, which may not be necessarily available onboard space crafts. In such cases, more dynamic and self-managing methods employing multi-agent techniques can be employed.

\begin{figure}
\centering
\includegraphics[width = 8cm]{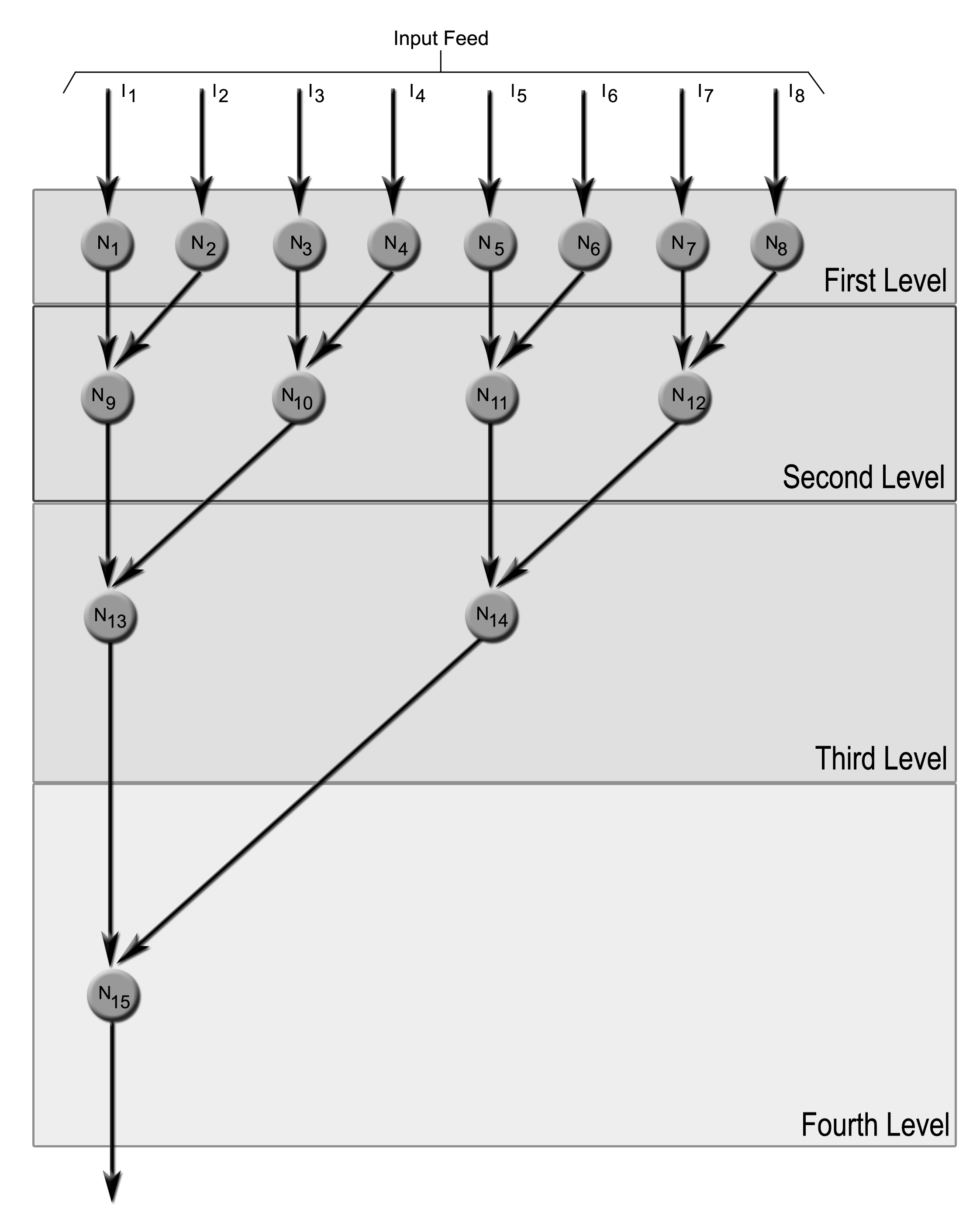}
\caption{Illustration of the Parallel Summation Algorithm}
\label{Figure 1}
\end{figure}

Keeping the above requirements in mind, parallel summation, which is an exemplar of parallel reduction algorithm is considered in this paper and illustrated in figure \ref{Figure 1}. The algorithm works in four sequential levels. The first level comprising nodes $N_{1} - N_{8}$ receives a live input feed of data. The second level comprising nodes $N_{9} - N_{12}$ receives data from the first level, adds the data received and yields the result to the third level nodes $N_{13}$ and $N_{14}$. The fourth level, adds data received from the third level nodes and produces the final result.

For a given time step, every node in a level operates in parallel. Each node is characterized by input dependencies (process or processor a node is dependent on for receiving an input), output dependencies (process or processor a node yields data to as output) and data contained in the node. The first level nodes have one input dependency and one output dependency. For instance, node $N_{1}$ has one input dependency $I_{1}$ and node $N_{9}$ as its output dependency. However, the second, third and fourth levels have two input dependencies and one output dependency. For instance, node $N_{13}$ of the third level has nodes $N_{9}$ and $N_{10}$ as input dependencies and node $N_{15}$ as output dependency. The data contained in a node is either the input data for the first level nodes or a calculated value (sum of two value in the case of a parallel summation algorithm) stored within a node.

The fault tolerance of the above parallel summation algorithm could be improved if in some way the algorithm itself could be self-managing. A simple definition of self-management in this context would be where if a node employed in the execution of the algorithm is about to fail, then the agent situated on the node predicted to fail can be moved off the node and the input and output dependencies re-established on another node. This would require the individual agents to incorporate intelligence whereby the condition of the computing node can be monitored and the agent moved if failure is predicted.

\begin{figure}
\centering
        \includegraphics[width=\textwidth]{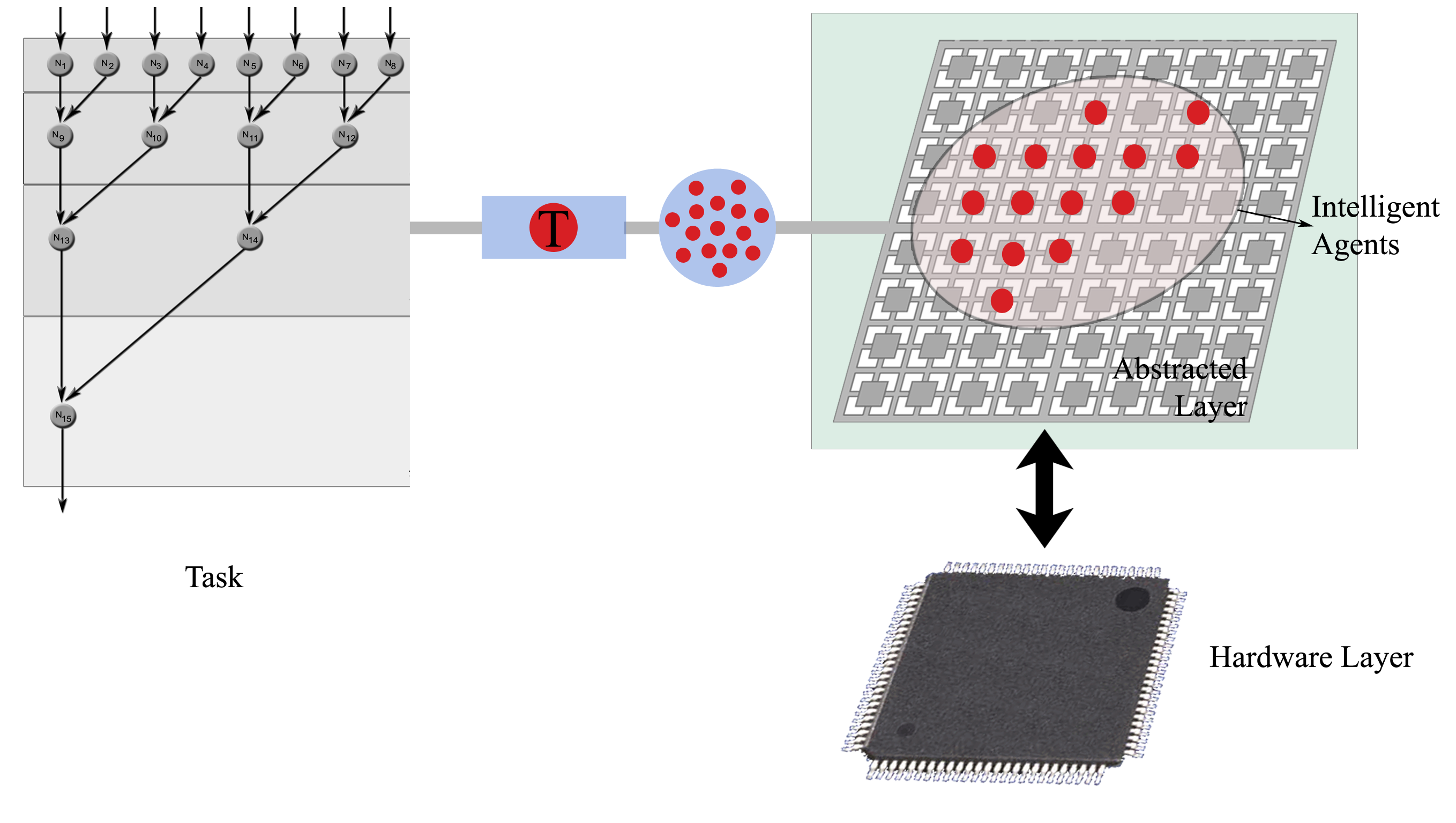}
\caption{Executing the parallel summation algorithm using the intelligent agent based approach}
\label{Figure99}
\end{figure}

To incorporate intelligence all the parallel components of the parallel summation algorithm shown in figure \ref{Figure99}, left, are mapped onto a set of agents such that the algorithm is essentially the payload of the agents. Figure \ref{Figure99}, middle, shows the agents. The set of agents then carry the payload onto the array of computing nodes. If a core failure is predicted by an agent, then the agent avoids the obstacle. Figure \ref{Figure99}, right, shows a set of agents that have located on the computing array.

\section{Experimental Studies with Intelligent Agents}
\emph{How can agent capabilities be implemented for fault tolerance?}

In this section, experimental studies To implement the intelligent agents considered in the above sections, it is necessary to firstly consider the requirements for the experiment, which include considerations of the computing platform and middleware. Then, the experiment performed for confirming the feasibility of the proposed approach is presented.

\subsection{Requirements}
A computer cluster-based parallel summation algorithm is considered in this section. The computing platform  was chosen as a cluster on arbitrary grounds and for two technical reasons. Firstly, a cluster is often characterized by three basic elements, namely a collection of nodes, a network connecting these nodes and a facility to access and share information between the nodes \cite{28}, which are sufficient constituents for providing an infrastructure for implementing intelligent agents. Secondly, existing middleware for clusters, namely Message Passing Interface (MPI) \cite{29} provide standard and portable programming interfaces.

The cluster used for the research reported in this paper is one among the high performance computing resources available at the Centre for Advanced Computing and Emerging Technologies (ACET), University of Reading, United Kingdom \cite{31} \cite{32}. The cluster is primarily used for the purpose of teaching and performing multi-disciplinary research. The cluster consists of a head node and 33 compute nodes. The formal specification of the head node is an Intel Pentium 4 CPU 3.20 GHz, 2 GB RAM and 160 GB hard disk, while that of 31 compute nodes are Intel Pentium 4 CPU 2.40 GHz, 512 MB RAM and 80 GB hard disk, and that of the remaining 2 compute nodes are Intel Pentium 4 CPU 2.60 GHz, 512 MB RAM and 40 GB hard disk. All nodes are connected via a Gigabit ethernet switch and communicate via the standard TCP protocol.

The cluster-based implementations reported in this paper are based on the Message Passing Interface (MPI), a standardized application programming interface (API) used for parallel and/or distributed computing. Open MPI \cite{33} \cite{34} version 1.3.3, an open source implementation of MPI 2.0 is employed on the cluster. An important feature of MPI 2.0, dynamic process creation and management, is of potential for exploration in the context of swarm-array computing.

The MPI dynamic process model permits the creation and management of a set of processes both when an MPI application begins and after the application has started. The management of newly created processes include cooperative termination of a process, communication between newly created processes and existing MPI application, and establishing communication between two independent processes. MPI\_COMM\_SPAWN is used to create a new MPI process and establish communication from an existing MPI application. On the other hand, MPI\_COMM\_ACCEPT and MPI\_COMM\_CONNECT can be used to establish communication between two independent processes. More MPI specific details on dynamic process model can be obtained from \cite{29} \cite{35}.

\subsection{Experiment}

The fault tolerant concepts incorporated using intelligent agents in the parallel summation algorithm are with respect to the cognitive capabilities presented in Section 3. Since MPI gives control over the process being executed rather than the processor on which a process is being executed, it was appropriate to implement the intelligent agent approach using MPI.

In order to implement the approach, an abstraction layer of the hardware resource layer had to be implemented. The hardware resource layer comprises physical nodes of the cluster and is connected via a switch, thereby forming a fully connected mesh topology. However, the abstracted layer is obtained when the physical nodes are abstracted as logical nodes. This is possible by implementing rules/policies. The policies are such that a process can only communicate with a vertically, horizontally or diagonally adjacent process, effectively leading to a grid topology on the abstracted layer. For example, nine nodes forming a fully connected mesh topology in figure 2 is abstracted to a grid topology in the abstraction layer.

\begin{figure}
\centering
\includegraphics[width = 10cm]{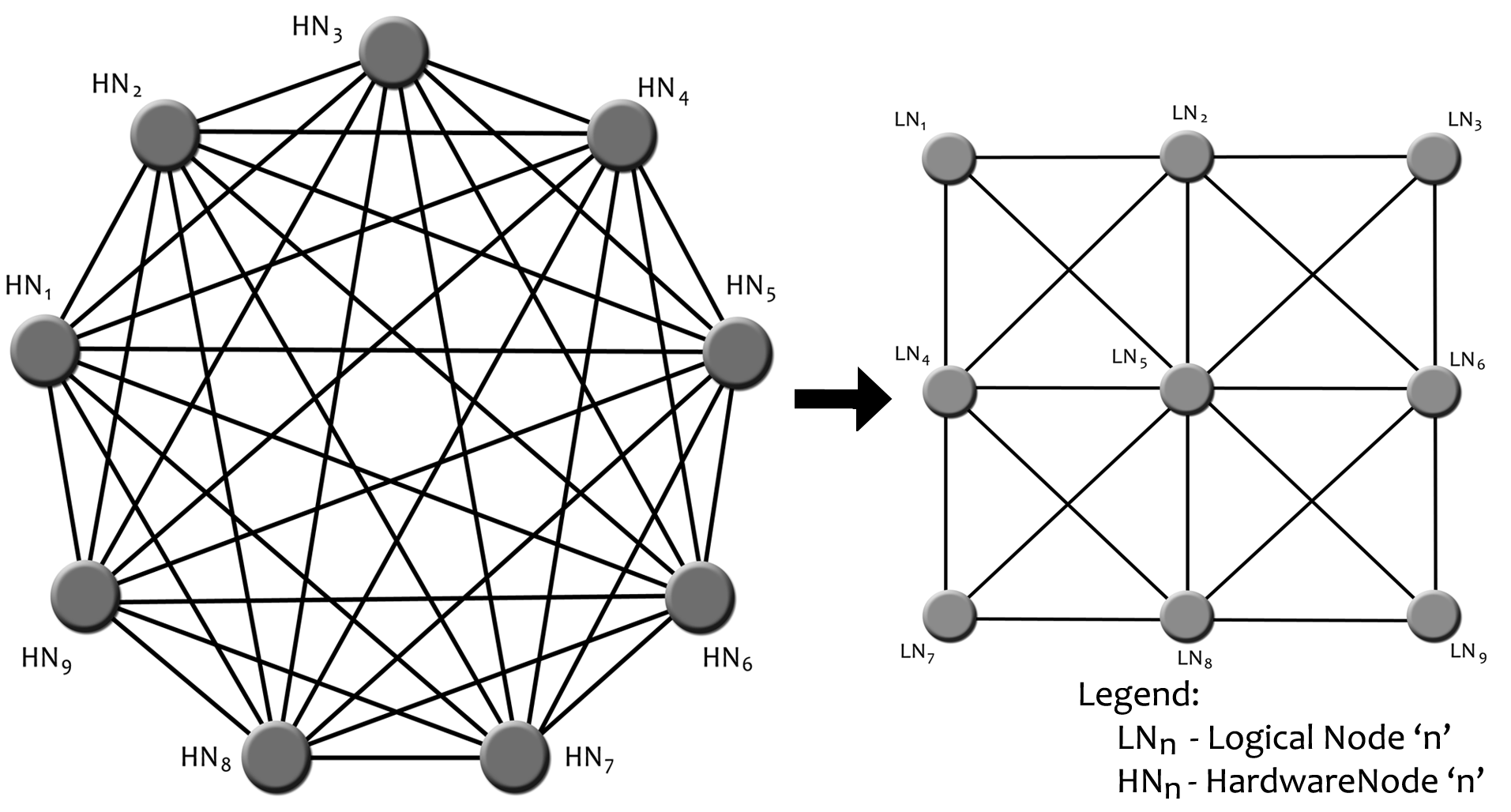}
\caption{Mapping hardware nodes to logical nodes}
\label{Figure 2}
\end{figure}

The agents on the abstracted layer are created such that they carry input and output dependencies and data. Since, parallel summation is relatively less complex when compared to other computational algorithms, the agents carry little information and have only few dependencies.

Each process executing on a node also gathers some sensory information through a hardware probing process to predict whether a node is likely to fail, on similar lines to proactive fault tolerance. The sensory information enables an agent to know its own surroundings on the computational environment, hence achieving the first cognitive capability considered in Section 3.

In the implementation presented in this paper node temperatures are simulated. When the temperature of a node rises beyond a threshold, the hardware probing process executing on that node predicts a failure and hence causes an agent to spawn a process on an adjacent core in the abstracted layer. In this case, an agent gathers sensory information on rising temperature than can likely impair or deteriorate its functioning, thereby achieving the third cognitive capability considered in Section 3. In the scenario considered in this paper, it is assumed that no adjacent core will fail in the next time step and hence any core adjacent to the core predicted to fail is randomly.

When rising temperature is detected, an agent identifies a node in the computational environment on which a new process can be spawned, thereby achieving the second cognitive capability considered in Section 3.

The agent on the abstracted core expected to fail shifts to the adjacent core on which the new process was spawned. An agent is capable of passing from one node to another, thereby achieving the fourth cognitive capability considered in Section 3.

The dependency information carried by the agent that was shifted to the new core is employed to reinstate the state of execution of the algorithm. The data for summation contained in the agent, either obtained from a previous level or a calculated value to be yielded to the next level, ensures that information is not lost and does not affect the final solution in critical applications.

\begin{figure}
\centering
\includegraphics[width = \textwidth]{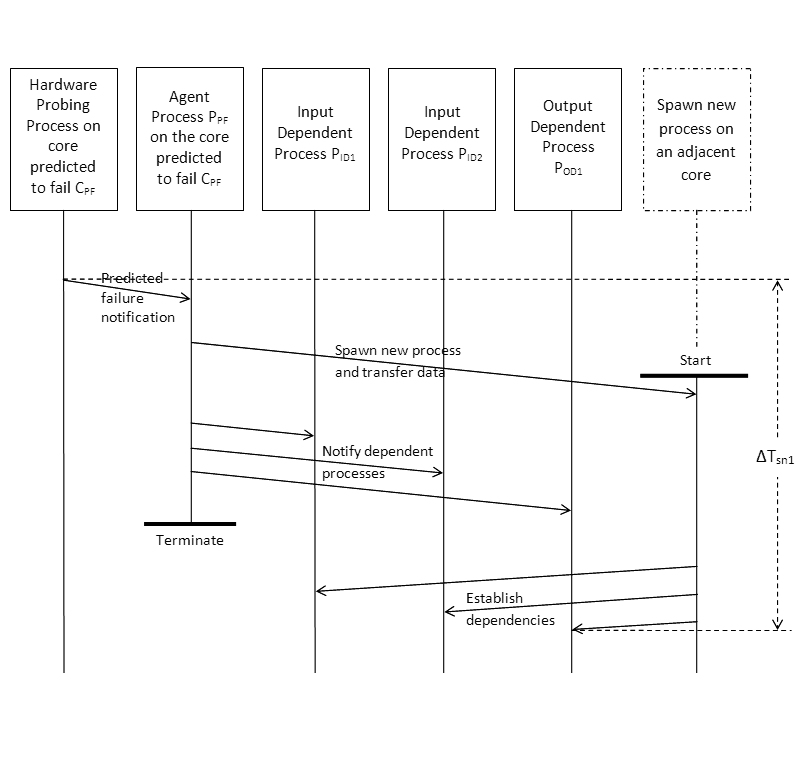}
\caption{Illustration of the Intelligent Agent Approach}
\label{Figure 3X}
\end{figure}

The above experiment is illustrated in figure \ref{Figure 3X} and briefly demonstrates how cognitive capabilities, namely perception, reasoning, judging, response and learning, can lead towards achieving fault tolerance. The approach implemented above is a simple demonstration that accommodates some of the concepts of intelligent agents and is a preliminary step towards realizing the approach.

\subsection{Results}
$T_{N_{n}}$, the time taken by an agent to transfer from a node $N_n$ predicted to fail onto an adjacent node in the abstracted layer and re-establish all process dependencies for seamless execution was noted. Nodes $N_{9} - N_{15}$ as shown in figure 1 are the computational nodes of the parallel summation algorithm, and hence are the only nodes considered for calculating $T_{N_{n}}$. Thirty different trial runs were performed to gather the statistic.

\begin{figure}
\centering
\includegraphics[width = 12.5cm]{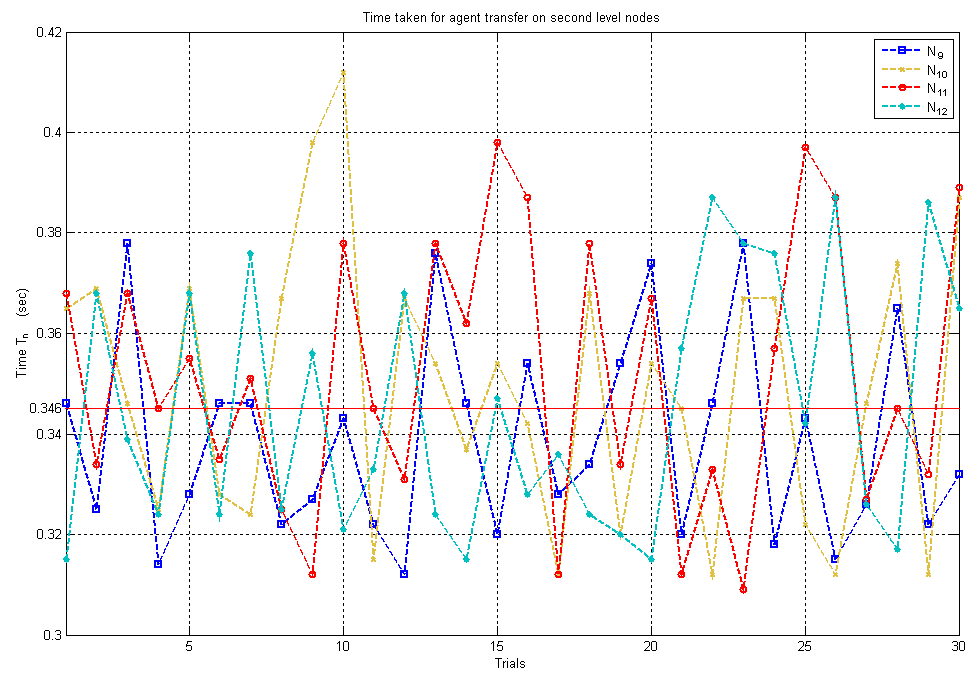}
\caption{Time taken for an agent transfer from a computational node in the second level to an adjacent node. Mean time for agent transfer in second level nodes $MT_{L_{2}} = 0.346 sec$}
\label{Figure 3}
\end{figure}

\begin{figure}
\centering
\includegraphics[width = 12.5cm]{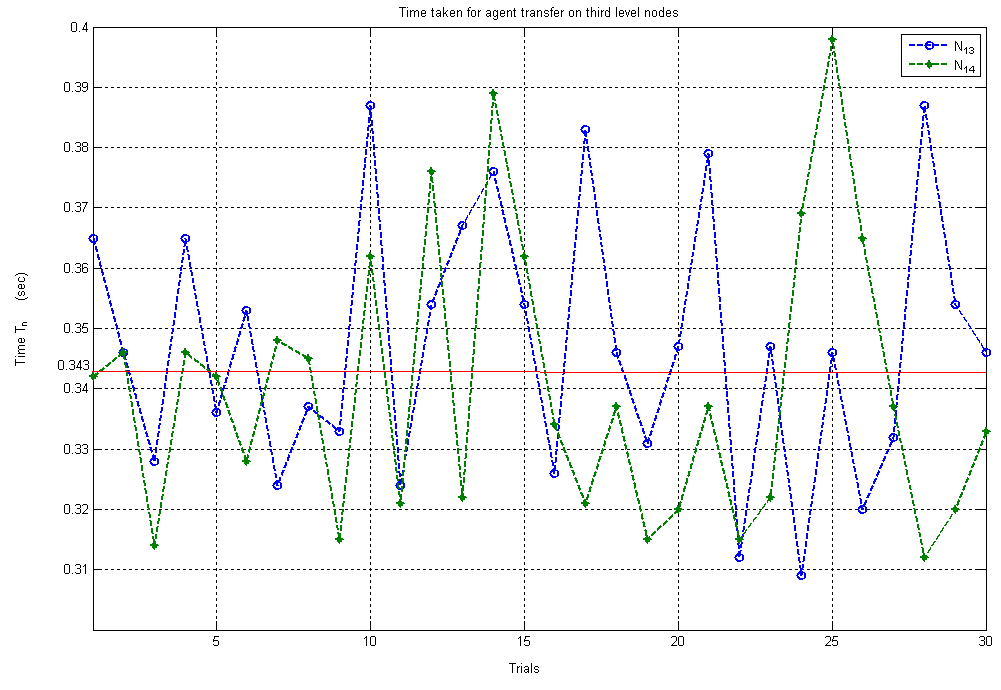}
\caption{Time taken for an agent transfer from a computational node in the third level to an adjacent node. Mean time for agent transfer in third level nodes $MT_{L_{3}} = 0.343 sec$}
\label{Figure 4}
\end{figure}

\begin{figure}
\centering
\includegraphics[width = 12.5cm]{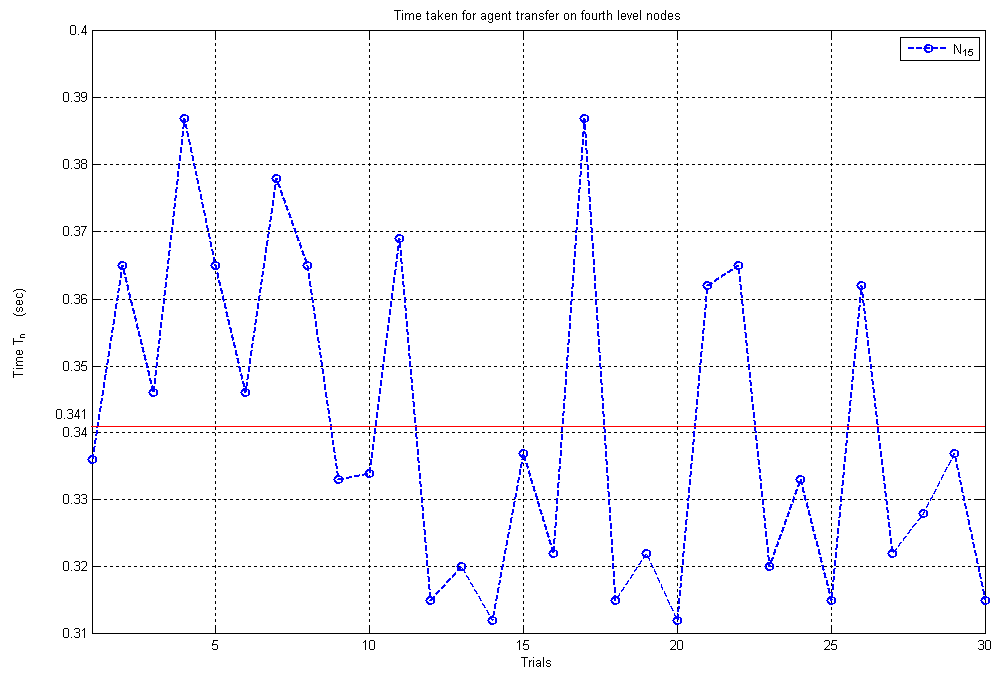}
\caption{Time taken for an agent transfer from a computational node in the fourth level to an adjacent node. Mean time for agent transfer in fourth level node $MT_{L_{4}} = 0.341 sec$}
\label{Figure 5}
\end{figure}

Figure \ref{Figure 3}, figure \ref{Figure 4} and figure \ref{Figure 5} are plots that show $T_{N_{n}}$ for 30 different trials. Figure 3 shows $T_{N_{n}}$ for the second level nodes $N_{9} - N_{12}$ for 30 trials. Figure 4 shows  $T_{N_{n}}$ for the third level nodes $N_{13}$ and $N_{14}$ for 30 trials. Figure 5 shows  $T_{N_{n}}$ for the fourth level node $N_{15}$ for 30 trials.

Further, $MT_{N_{n}}$, the mean time of $T_{N_{n}}$ for a particular node was calculated. This metric yields information on the mean time taken by an agent to transfer from a node $N_{n}$ predicted to fail onto an adjacent node in the abstracted layer and re-establish all process dependencies for seamless execution. $MT_{N_{n}}$ is calculated as $MT_{N_{n}} = \left( \sum_{TR=1}^{6} T_{n} \right) \bigg/ 30, n = 9 \cdots 15$.

$MT_{L_{p}}, p = 1, 2, 3$, the mean time taken for an agent transfer from all nodes predicted to fail in a level of the parallel summation algorithm onto an adjacent node in the abstracted layer was calculated. Nodes $N_{9} - N_{12}$ are used in level 2, while $N_{13}$ and $N_{14}$ in level 3 and $N_{15}$ in level 4. $MT_{L_{p}}$ is calculated as $MT_{L_{2}} = \left( \sum_{n = 9}^{12} MT_{N_{n}} \right) \bigg/ 4$, $MT_{L_{3}} = \left( \sum_{n = 13}^{14} MT_{N_{n}} \right) \bigg/ 2$ and $MT_{L_{4}} = MT_{N_{15}}$.

The mean time for an agent transfer from a computational node in the second level to an adjacent node in the abstracted layer is obtained as $MT_{L_{2}} = 0.346 sec$, indicated by an axis line in figure 3. The mean time for an agent transfer from a computational node in the third level to an adjacent node in the abstracted layer is obtained as $MT_{L_{3}} = 0.343 sec$, indicated by an axis line in figure 4. The mean time for an agent transfer from a computational node in the fourth level to an adjacent node in the abstracted layer is obtained as $MT_{L_{4}} = 0.341 sec$, indicated by an axis line in figure 5.

$MT_{N_{N}}$, the mean time of agent transfer for all computational nodes in the parallel summation algorithm onto an adjacent node in the abstracted layer was calculated. This value can be calculated as the mean time of all $MT_{N_{n}}$ of the computational nodes or the mean time of all $MT_{L_{p}}$ of the computational levels. $MT_{N_{N}}$ is calculated as $MT_{N_{N}} = \left( \sum_{n=9}^{15} MT_{N_{n}} \right) \bigg/ 7$ or
$MT_{N_{N}} = \left( \sum_{p=2}^{4} MT_{L_{p}} \right) \bigg/ 3$.


\vspace*{13pt}
\centerline{\footnotesize Table~1. Computed values for $MT_{T_{n}}$, $MT_{L_{p}}$ and $MT_{N_{N}}$, $n = 9 \cdots 15$, $p = 2, 3, 4$.}
\begin{center}
\begin{tabular}{ c | c | c | c | c | c | c | c }
\hline
$p$                 &\multicolumn{4}{| c |}{2}  &\multicolumn{2}{| c |}{3}  &4\\
\hline
$n$                 &9      &10     &11     &12     &13     &14     &15\\
\hline
\hline
$MT_{T_{n}}$ \footnotesize{(sec)}       &{0.339}     &{0.349}     &{0.352}     &{0.345}     &{0.347}     &{0.340}     &{0.341}\\
\hline
$MT_{L_{p}}$ \footnotesize{(sec)}       &\multicolumn{4}{| c |}{0.346}     &\multicolumn{2}{| c |}{0.343}     &{0.341}\\
\hline
$MT_{N_{N}}$ \footnotesize{(sec)}       &\multicolumn{7}{| c }{0.344}\\
\hline
\end{tabular}
\end{center}

\vspace{8pt}

Table 1 summaries the computed values of $MT_{T_{n}}$, $MT_{L_{2}}$, $MT_{L_{3}}$, $MT_{L_{4}}$ and $MT_{N_{N}}$. The mean time $MT_{N_{N}}$ of all the computational nodes in the agent-based approach is calculated as 0.344 sec. This value is $\Delta T_{sn_{1}}$ as shown in figure \ref{Figure 3X}, the time taken for reinstating execution after a predicted node failure. If traditional checkpointing with human adminstration was employed, reinstating execution would be at least in the order of minutes. This brief comparison reveals that the multi-agent approach is more effective than traditional fault tolerant methods.

\begin{figure}
\centering
\includegraphics[width = 12.5cm]{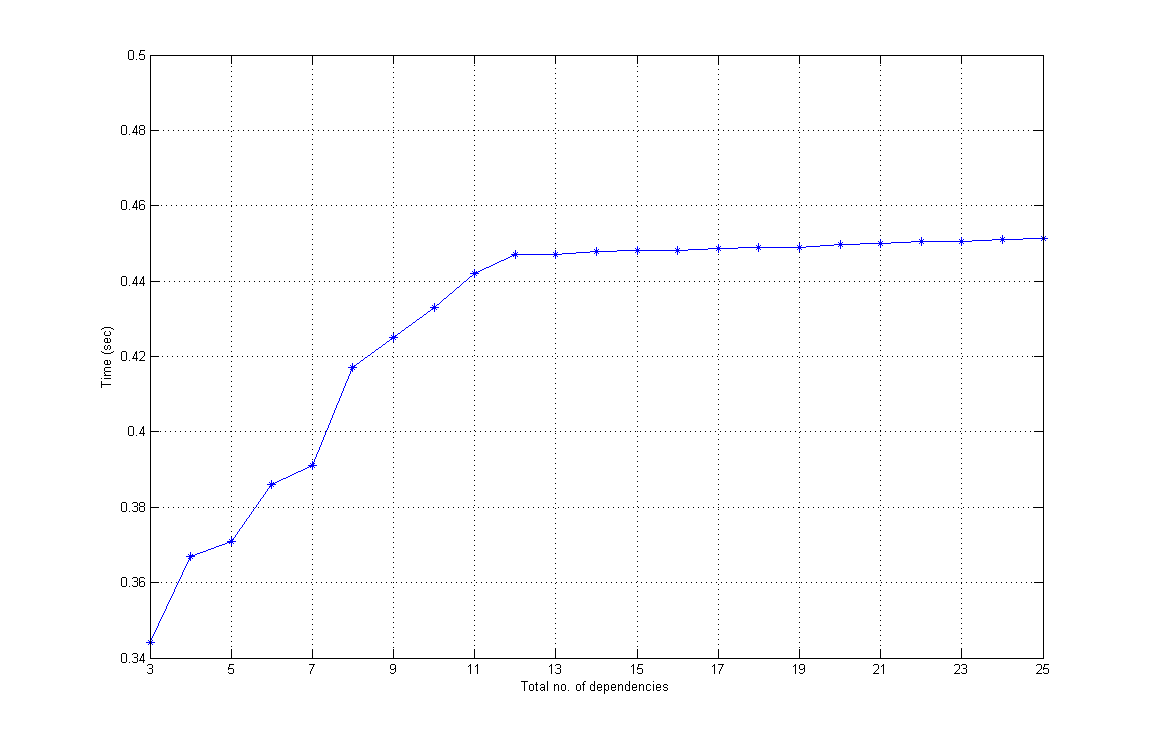}
\caption{Graph plotted for the Number of dependencies vs. Mean time for reinstating execution in the intelligent agent approach.}
\label{Figure 6}
\end{figure}

It is also worthwhile to consider how the time taken by an agent to reinstate execution would be affected by increasing dependencies (Total dependencies being equal to the sum of the input and output dependencies. In the experiments presented in this paper each node had only one output dependency). For this, experiments were conducted such that the mean time $MT_{N_{N}}$ was calculated for the parallel summation algorithm with different input dependencies. The graph shown in figure \ref{Figure 6} is based on the results obtained from the experiments.

The general trend of the graph is such that there is an increase in the time taken for reinstating execution until there are twelve dependencies. This is due to the fact that there is an increase in the overheads associated. However, the graph is steady for dependencies greater than twelve. In other words, there is little increase in the mean time for reinstating execution though the algorithm starts to handle more data.

In short, though only preliminary results obtained through simple experiments are presented in this paper, the proposed multi-agent approach is promising and paves a path for effectively achieving fault tolerance in parallel computing systems.

\section{Conclusion}

In this paper the incapability of traditional methods such as checkpointing for achieving efficient and effective fault tolerance in complex parallel computing systems is presented. Hence, the need for a transition from traditional checkpointing to agent-based methods is highlighted. To extend agent-based methods for large scale parallel computing systems three fundamental questions need to be addressed. In this regard, ``What agent capabilities are required for fault tolerance?'', ``What parallel computational tasks can benefit from such agent capabilities?'' and ``How can agent capabilities be implemented for fault tolerance?'' are addressed.

Cognitive capabilities of agents on a computing environment that can lead to fault tolerance are presented. Parallel reduction algorithms are considered as tasks that can benefit from such agent capabilities. An agent-based approach is implemented on a computer cluster using the Message Passing Interface (MPI). Experimental results are gathered based on the time taken for reinstating execution once a fault is predicted to occur. Though approximations and assumptions are made in the experiments, preliminary results confirm that the proposed method, if well implemented, is more beneficial and dependable when compared to traditional methods. Hence this paper is an effort towards realising an implementation that can be employed to achieve improvised fault tolerance and confirms that \emph {agent intelligence can be used to achieve fault tolerant parallel computing systems}.

Future work will aim to analyse the method using metrics that can evaluate various aspects of the method such as precision in fault prediction, capability to prevent faults and reduction of overhead in recovery from faults. Immediate efforts will be made to address real-time issues in the implementation by considering multiple node failures as against single node failures reported in this paper. The approach will also be considered for being implemented and tested on other parallel computing environments. Further, a more sophisticated and general implementation of the approach proposed above will be considered.


\end{document}